\documentclass[preprint,prd,aps,showpacs,showkeys,nofootinbib]{revtex4}

\usepackage{graphicx}
\usepackage{amsmath}
\usepackage{amssymb}

\begin{document}

\title {\large The c-quark EDM and production of $h_c$ in
$e^+e^-$ annihilation}

\author{Xiao-Jun Bi$^a$, Tai-Fu Feng$^b$, Xue-Qian Li$^c$, Jukka Maalampi$^b$, Xinmin Zhang$^d$}
\affiliation{$^a$Key Laboratory of Particle Physics, Institute of High Energy Physics, Chinese Academy of
Sciences, P.O. Box 918-3, Beijing 100049, People's Republic of China\\\\
$^b$Department of Physics, 40014 University of Jyv\"askyl\"a,Finland\\\\
$^c$Department of Physics, Nankai University, Tianjing 300071, P. R. China\\\\
$^d$Theoretical Physics Division,  Institute of High Energy Physics, Chinese Academy of
Sciences, P.O. Box 918-4, Beijing 100049, People's Republic of China}

\date{\today}

\begin{abstract}

We  analyze the charm quark electric dipole moment (EDM) in the
minimal supersymmetric extension of the standard model (MSSM),
including important two-loop gluino contributions. Considering the
experimental constraint on the neutron EDM, the theoretical
prediction for the charm quark EDM can reach about $10^{-20}e\cdot
cm$. If taking into account the mixing between the scharm and stop
quarks in the effective supersymmetry scenario, the charm quark
EDM can be enhanced to $\sim 10^{-19}e\cdot cm$. Direct production
of the CP-odd $^1P_1$ state $h_c$ in $e^+e^-$ annihilation via the
CP-violating process at the BES-III and CLEO-C is analyzed.

\end{abstract}

\pacs{11.30.Er, 12.60.Jv,14.80.Cp}

\keywords{electric dipole moment, charm quark, supersymmetry}

\maketitle

\section {introduction}

Up to now, CP violation is only found in the $K$-\cite{cpk} and
$B$-system\cite{cpb}, which can be well explained within the
standard model (SM) of electroweak interaction. As is well known,
electric dipole moment (EDM) of an elementary particle is a clear
signature of CP violation. However, EDM of a fermion does not
appear up to two-loop order, and the three-loop contributions
partially chancel among each others in the SM. Therefore,
observation of sizable EDM of an elementary fermion would
definitely be a signal of existence of new physics beyond the SM
with extra CP phases.

Supersymmetry (SUSY) is now believed to be the most attractive
scenario for new physics. Beside the Cabibbo-Kobayashi-Maskawa
(CKM) mechanism, the soft breaking terms provide a new source of
$CP$ and flavor violation in the minimal supersymmetric extension
of the standard model (MSSM). Those $CP$ violating phases can
affect some important observables in the mixing of Higgs bosons
\cite{higgs}, the lepton and neutron electric dipole moments
(EDMs) \cite{edm1, edm2}, lepton polarization asymmetries in the
semi-leptonic decays \cite{semi-lep}, the production of $P$-wave
charmonium and bottomonium \cite{quarknium}, and $CP$ violation in
rare $B$-decays and $B^0\bar{B}^0$ mixing \cite{dmix}. At present,
the strictest constraints on those $CP$ violation phases originate
from the lepton and neutron EDMs.
Since the lepton and neutron EDMs have not been measured so far,
there are several suggestions to keep the theoretical estimate of
the neutron and electron EDMs below the experimental limits, they
are (i) choosing small CP phases $\lesssim
\mathcal{O}(10^{-3})$\cite{phase}, (ii) finding appropriate
parameter domain where various contributions cancel with each
other\cite{cancel}, or (iii) making the first two generations of
scalar fermions heavy enough (heavier than 20 TeV) while keeping
the soft masses of the third generation below TeV to keep the
Higgs boson naturally light\cite{heavy}.

In this work we calculate the charm quark EDM in the $CP$
violating MSSM. As pointed out in \cite{edm1}, the quark
chromoelectric dipole moment (CEDM) makes important contributions
to the quark EDM at a low energy scale. In addition, there is a
parameter space where some two-loop gluino diagrams provide
important contributions to the quark EDM. In our analysis, we
include all the contributions from those pieces. We first work in
the simplest model, neglecting the new possible flavor violation
sources except the CKM mechanism, and avoiding any assumptions
about unification of the soft breaking parameters. We find that
the charm EDM can reach about $10^{-20} e\cdot cm$. Then we also
calculate the charm quark EDM in the effective supersymmetry
scenario\cite{eff} where the scenario (iii) mentioned above is
accommodated. Since the first two-generation sfermions are very
heavy, the charm quark EDM is greatly suppressed in this scenario.
However, while considering large mixing between the second and the
third generation sfermions we find that the charm quark EDM can be
greatly enhanced to as large as about $10^{-19} e\cdot cm$. The
reason is that, since the effective operator which determines the
charm quark EDM induces a chiral flip, the large mixing between
the scharm and stop can enhance the charm quark EDM due to the top
quark mass.

The analysis in Refs. \cite{cedm,bedm} shows that the EDM of
heavy quark plays an important role in the direct production of
singlet P-wave quarkonia, thus with upgraded BEPC\cite{bepc} and
CLEO-C programs\cite{cleo}, it might be possible to produce the
$^1$P$_1$ charmonium state, $h_c$, directly. Conversely, if no
$h_c$ meson is observed, an upper bound on the charm quark EDM
would be set.

The paper is organized as follows: In the next section, we give
the analytic expressions of the charm quark EDM and CEDM in $CP$
violating MSSM. In Section III we present the numerical results.
Section IV is devoted to discuss  $h_c$ production which is
closely associated with the magnitude of the charm quark EDM, at
$e^+e^-$ colliders. Finally we conclude in section V.

\section{The charm quark EDM}

\begin{figure}[t]
\setlength{\unitlength}{1mm}
\begin{center}
\begin{picture}(0,20)(0,0)
\put(-62,-100){\includegraphics{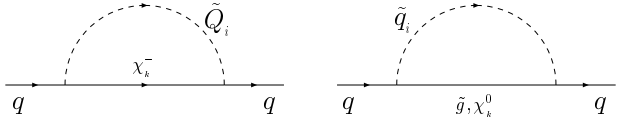}}
\end{picture}
\caption[]{The one-loop self energy diagrams which lead to quark
EDM and CEDM in the MSSM, the corresponding triangle diagrams are
obtained by attaching a photon or  gluon line to the self-energy
blob in all possible ways.} \label{fig1}
\end{center}
\end{figure}

\indent\indent In the effective Lagrangian, the charm quark EDM is
defined through a dimension-five operator
\begin{eqnarray}
&&{\cal L}_{_{EDM}}=-\frac{i}{2}d_{_c}\overline{c}\sigma^{\mu\nu}\gamma_5
cF_{_{\mu\nu}}
\label{eq1}
\end{eqnarray}
with $F_{_{\mu\nu}}$ being the electromagnetic field strength. In
a theoretical framework with $CP$ violation, the corresponding
loop diagrams induce the charm EDMs. Since quarks also take part
in the strong interaction, the chromoelectric dipole moment (CEDM)
operator is $$\overline{c}T^a\sigma^{\mu\nu}\gamma_5
cG^a_{_{\mu\nu}},$$ where $T^a\;(a=1,\;\cdots,\;8)$ denote the
generators of the strong $SU(3)$ gauge group and $G_{\mu\nu}^a$ is
the chromo-electromagnetic field strength. This operator
contributes to $d_{_c}$ at low energy scale. In principle, the
dimension-six Weinberg operator and two-loop Bar-Zee diagrams all
provide contributions to $d_{_c}$. Nevertheless, the contributions
from the Weinberg operator and Bar-Zee diagrams to $d_{_c}$ can be
ignored safely as we consider the constraint on the neutron EDM.
The best way to describe certain loop-induced contributions is the
effective theory approach, where the heavy particles are
integrated out at the matching scale and the effective theory
includes a full set of $CP$ violating operators. In this work, we
restrict ourselves to the following operators that are relevant to
the c-quark EDM
\begin{eqnarray}
&&{\cal L}_{_{eff}}=-{i\over2}d_{_c}^\gamma\overline{c}\sigma^{\mu\nu}\gamma_5
cF_{_{\mu\nu}}-{i\over2}d_{_c}^g\overline{c}T^a\sigma^{\mu\nu}\gamma_5
cG_{_{\mu\nu}}^a\;.
\label{eq2}
\end{eqnarray}

The one-loop supersymmetric contributions to the Wilson
coefficients in Eq. (\ref{eq2}) originate from three types of
graphs: gluino-squark, chargino-squark, and neutralino-squark
loops (FIG.\ref{fig1}). Details about the descriptions are
presented in ref.\cite{Feng}. The contributions of the one-loop
gluino-squark diagrams are
\begin{eqnarray}
&&d_{_{\tilde{g}(1)}}^\gamma=-{2\over3\pi}e_{_u}e\alpha_{_s}\sum\limits
_{i=1}^2{\bf Im}\Big(({\cal Z}_{_{\tilde c}})_{_{2,i}}
({\cal Z}_{_{\tilde c}}^\dagger)_{_{i,1}}e^{-i\theta_{_3}}\Big)
\nonumber\\
&&\hspace{1.2cm}\times
{|m_{_{\tilde g}}|\over m_{_{\tilde{c}_i}}^2}B\Big({|m_{_{\tilde g}}|^2
\over m_{_{\tilde{c}_i}}^2}\Big)
\;,\nonumber\\
&&d_{_{\tilde{g}(1)}}^g={g_3\alpha_{_s}\over4\pi}\sum\limits
_{i=1}^2{\bf Im}\Big(({\cal Z}_{_{\tilde c}})_{_{2,i}}
({\cal Z}_{_{\tilde c}}^\dagger)_{_{i,1}}e^{-i\theta_{_3}}\Big)
\nonumber\\
&&\hspace{1.2cm}\times
{|m_{_{\tilde g}}|\over m_{_{\tilde{c}_i}}^2}C\Big({|m_{_{\tilde g}}|^2
\over m_{_{\tilde{c}_i}}^2}\Big)\;.
\label{eqa3-1}
\end{eqnarray}
Here $\alpha=e^2/(4\pi)$, $\theta_{_3}$ denotes the phase of $m_{_{\tilde g}}$, and ${\cal
Z}_{_{\tilde q}}\;(q=u,\cdots,b)$ are the mixing matrices of
squarks, i.e. ${\cal Z}_{_{\tilde q}}^\dagger{\bf m}_{_{\tilde
q}}^2{\cal Z}_{_{\tilde q}} =diag(m_{_{{\tilde
q}_1}}^2,\;m_{_{{\tilde q}_2}}^2)$ where
\begin{widetext}
\begin{eqnarray}
&&{\bf m}_{_{\tilde q}}^2=\left(\begin{array}{cc}
m_{_{\tilde Q}}^2+m_{_q}^2+m_{_{\rm z}}^2({1\over2}-Q_{_q}s_{_{\rm w}}^2)
\cos2\beta&m_{_q}(A_{_q}^*-\mu R_{_q})\\m_{_q}(A_{_q}-\mu^*R_{_q})&
m_{_{\{{\tilde U},{\tilde D}\}}}^2+m_{_q}^2+m_{_{\rm z}}^2({1\over2}-Q_{_q}s_{_{\rm w}}^2)
\cos2\beta\end{array}\right)\;,
\label{eqa3-a}
\end{eqnarray}
\end{widetext}
with $Q_{_{q}}=2/3(-1/3),\;R_{_q}=\tan\beta(1/\tan\beta)$ for
$q=u\;(d)$. $\tan\beta={\upsilon_{_u}\over\upsilon_{_d}}$ is the
ratio between the up- and down-type Higgs vacua, and
$\theta_{_{\rm w}}$ is the Weinberg angle. The shortened notations
$s_{_{\rm w}}=\sin\theta_{_{\rm w}},\;c_{_{\rm
w}}=\cos\theta_{_{\rm w}}$ are adopted. The loop functions are
$$B(r)=[2(r-1)^2]^{-1} [1+r+2r\ln
r/(r-1)],\;C(r)=[6(r-1)^2]^{-1}[10r-26-(2r-18)\ln r/(r-1)].$$ In a
similar way, the one loop neutralino-squark contributions can be
written as
\begin{eqnarray}
&&d_{_{\chi_{_k}^0(1)}}^\gamma=e_{_u}{e\alpha\over16\pi s_{_{\rm w}}^2
c_{_{\rm w}}^2}\sum\limits_{i,k}{\bf Im}\Big((A_{_N}^c)_{_{k,i}}
(B_{_N}^c)^\dagger_{_{i,k}}\Big)
\nonumber\\
&&\hspace{1.2cm}\times
{m_{_{\chi_{_0}^k}}\over m_{_{\tilde{c}_i}}^2}B\Big({m_{_{\chi_{_0}^k}}^2
\over m_{_{\tilde{c}_i}}^2}\Big)
\;,\nonumber\\
&&d_{_{\chi_{_k}^0(1)}}^g={g_3\alpha_{_s}\over64\pi s_{_{\rm w}}^2
c_{_{\rm w}}^2}\sum\limits_{i,k}{\bf Im}\Big((A_{_N}^c)_{_{k,i}}
(B_{_N}^c)^\dagger_{_{i,k}}\Big)
\nonumber\\
&&\hspace{1.2cm}\times
{m_{_{\chi_{_0}^k}}\over m_{_{\tilde{c}_i}}^2}B\Big({m_{_{\chi_{_0}^k}}^2
\over m_{_{\tilde{c}_i}}^2}\Big)
\label{eqa3-2}
\end{eqnarray}
with
\begin{eqnarray}
&&(A_{_N}^c)_{_{k,i}}=-{4\over3}s_{_{\rm w}}({\cal Z}_{_{\tilde c}})_{_{2,i}}
({\cal Z}_{_N})_{_{1,k}}+{m_{_c}c_{_{\rm w}}\over m_{_{\rm w}}s_{_\beta}}
\nonumber\\
&&\hspace{1.8cm}\times
({\cal Z}_{_{\tilde c}})_{_{1,i}}({\cal Z}_{_N})_{_{4,k}}
\;,\nonumber\\
&&(B_{_N}^c)_{_{k,i}}=({\cal Z}_{_{\tilde c}})_{_{1,i}}\Big({s_{_{\rm w}}\over3}
({\cal Z}_{_N})_{_{1,k}}^*+c_{_{\rm w}}({\cal Z}_{_N})_{_{2,k}}^*\Big)
\nonumber\\
&&\hspace{1.8cm}
+{m_{_c}c_{_{\rm w}}\over m_{_{\rm w}}s_{_\beta}}
({\cal Z}_{_{\tilde c}})_{_{2,i}}({\cal Z}_{_N})_{_{4,k}}^*\;.
\label{eqa3-3}
\end{eqnarray}
Here $\alpha_{_s}=g_3^2/(4\pi)$, $s_{_\beta}=\sin\beta,
\;c_{_\beta}=\cos\beta$, and ${\cal Z}_{_N}$ is the mixing matrix 
of neutralinos.

\begin{figure}[b]
\setlength{\unitlength}{1mm}
\begin{center}
\begin{picture}(0,120)(0,0)
\put(-80,-80){\includegraphics{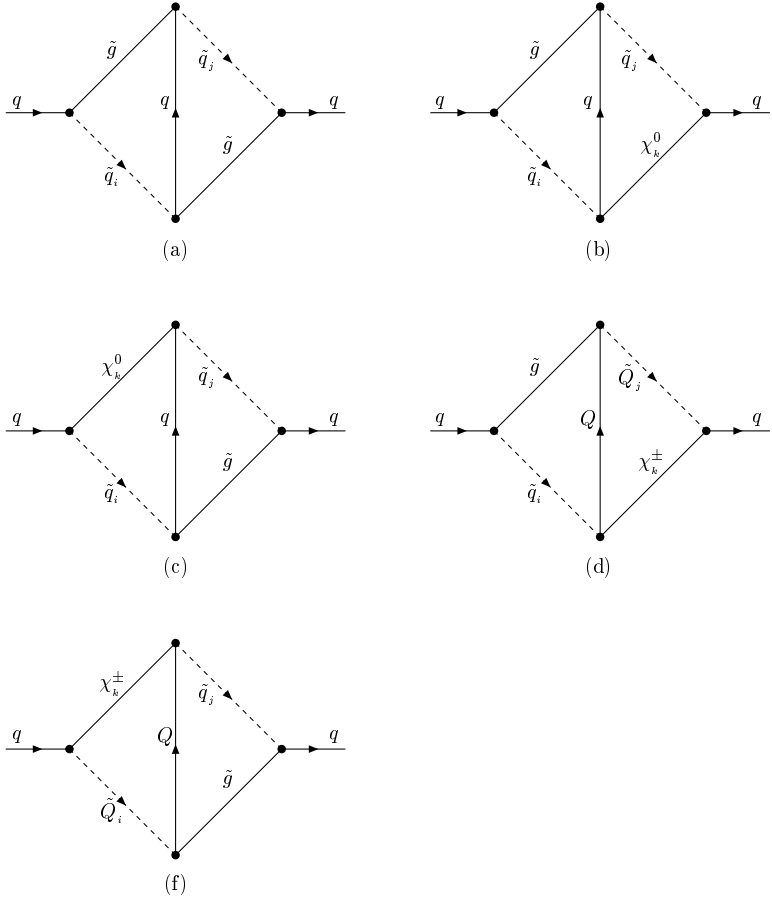}}
\end{picture}
\caption[]{The two-loop self energy diagrams which lead to quark
($q=u\;(d),\; Q=d\;(u)$) EDM and CEDM in the MSSM, the corresponding 
triangle diagrams are
obtained by attaching an external photon or  gluon line to the
self-energy diagrams in all possible ways.} \label{fig2}
\end{center}
\end{figure}

Finally, the chargino-squark contributions are formulated as
\begin{eqnarray}
&&d_{_{\chi_{_k}^\pm(1)}}^\gamma={e\alpha\over4\pi s_{_{\rm w}}^2}
V_{_{cQ}}^\dagger V_{_{Qc}}
\sum\limits_{i,k}{\bf Im}\Big((A_{_C}^Q)_{_{k,i}}(B_{_C}^Q)^\dagger
_{_{i,k}}\Big){m_{_{\chi_{_k}^\pm}}\over m_{_{\tilde{Q}_i}}^2}
\nonumber\\
&&\hspace{1.2cm}\times
\Big[e_{_d}B\Big({m_{_{\chi_{_k}^\pm}}^2\over m_{_{\tilde{Q}_i}}^2}\Big)
+(e_{_u}-e_{_d})A\Big({m_{_{\chi_{_k}^\pm}}^2\over
m_{_{\tilde{Q}_i}}^2}\Big)\Big]
\;,\nonumber\\
&&d_{_{\chi_{_k}^\pm(1)}}^g={g_3\alpha\over4\pi s_{_{\rm w}}^2}
V_{_{cQ}}^\dagger V_{_{Qc}}
\sum\limits_{i,k}{\bf Im}\Big((A_{_C}^Q)_{_{k,i}}(B_{_C}^Q)^\dagger
_{_{i,k}}\Big)
\nonumber\\
&&\hspace{1.2cm}\times
{m_{_{\chi_{_k}^\pm}}\over m_{_{\tilde{Q}_i}}^2}
B\Big({m_{_{\chi_{_k}^\pm}}^2\over m_{_{\tilde{Q}_i}}^2}\Big)
\label{eqa3-4}
\end{eqnarray}
where $V$ denotes the CKM matrix, and the loop function is
$$A(r)=2(1-r)^{-2}[3-r+2\ln r/(1-r)].$$ The couplings are defined as
\begin{eqnarray}
&&(A_{_C}^d)_{_{k,i}}={m_{_u}\over\sqrt{2}m_{_{\rm w}}s_{_\beta}}
({\cal Z}_{_{\tilde d}})_{_{1,i}}({\cal Z}_+)_{_{2,k}}
\;,\nonumber\\
&&(B_{_C}^d)_{_{k,i}}={m_{_d}\over\sqrt{2}m_{_{\rm w}}c_{_\beta}}
({\cal Z}_{_{\tilde d}})_{_{2,i}}({\cal Z}_-)_{_{2,k}}
-({\cal Z}_{_{\tilde d}})_{_{1,i}}({\cal Z}_-)_{_{1,k}}
\;,\nonumber\\
&&(A_{_C}^u)_{_{k,i}}={m_{_d}\over\sqrt{2}m_{_{\rm w}}c_{_\beta}}
({\cal Z}_{_{\tilde u}})_{_{1,i}}({\cal Z}_-)_{_{2,k}}^*
\;,\nonumber\\
&&(B_{_C}^u)_{_{k,i}}={m_{_u}\over\sqrt{2}m_{_{\rm w}}s_{_\beta}}
({\cal Z}_{_{\tilde u}})_{_{2,i}}({\cal Z}_+)_{_{2,k}}^*
-({\cal Z}_{_{\tilde u}})_{_{1,i}}({\cal Z}_+)_{_{1,k}}^*
\;,\label{eqa3-5}
\end{eqnarray}
where ${\cal Z}_\pm$ are the right- and left-handed mixing
matrices of the charginos. Noting that one-loop chargino
contributions to the quark EDMs and CEDMs are proportional to a
suppression factor ${m_{_q}\over m_{_{\rm w}}}$ which exists in
all the couplings $A_{_C}^q \;(q=u,\;d)$.

The two-loop gluino contributions originate from the two-loop self
energy diagrams (FIG. \ref{fig2}). The corresponding triangle
diagrams are obtained by attaching an external gluon or photon
line to the self-energy diagrams in all possible ways. In those
two-loop diagrams, there are no new suppression factors except the
loop integration factor in the corresponding amplitude, and the
resulting Wilson coefficients of those dipole moment operators do
not involve ultra-violet divergence. Furthermore, there is a
parameter space where the two-loop results are comparable with
those one-loop contributions because the dependence of the
two-loop results on the relevant $CP$ phases differs from that the
one-loop results on the corresponding $CP$ phases. The
contributions of the two-loop gluino-gluino diagrams  to the quark
EDMs and CEDMs are formulated as
\begin{eqnarray}
&&d_{_{\tilde{g}(2)}}^\gamma={8e_{_q}e\alpha_{_s}^2|m_{_{\tilde g}}|
\over9(4\pi)^2m_{_{\rm w}}^2}F_1(x_{_q}, x_{_{\tilde{q}_j}}
,x_{_{\tilde g}},x_{_{\tilde g}},x_{_{\tilde{q}_i}})
{\bf Im}\Big(({\cal Z}_{_{\tilde q}})_{_{2,j}}
({\cal Z}_{_{\tilde q}}^\dagger)_{_{j,1}}e^{-i\theta_{_3}}\Big)
\;,\nonumber\\
&&d_{_{\tilde{g}(2)}}^g={8g_3\alpha_{_s}^2|m_{_{\tilde g}}|\over9(4\pi)^2m_{_{\rm w}}^2}
F_3(x_{_q}, x_{_{\tilde{q}_j}},x_{_{\tilde g}},x_{_{\tilde g}},x_{_{\tilde{q}_i}})
{\bf Im}\Big(({\cal Z}_{_{\tilde q}})_{_{2,j}}({\cal Z}_{_{\tilde q}}^\dagger)_{_{j,1}}
e^{-i\theta_{_3}}\Big)\;,
\label{eq8}
\end{eqnarray}
where $x_i=m_i^2/m_{_{\rm w}}^2$, and the function $F_i(x_0,x_1
,x_2,x_3,x_4)\;(i=1,\;3)$ are defined in Ref.\cite{Feng}.

The contributions of the two loop neutralino-gluino  to the quark
EDM and CEDM are given by
\begin{eqnarray}
&&d_{_{\chi_{_k}^0(2)}}^\gamma=
{4e_{_q}e\alpha\alpha_{_s}\over3(4\pi)^2s_{_{\rm w}}^2c_{_{\rm w}}^2
m_{_{\rm w}}^2}\Big\{|m_{_{\tilde g}}|F_1(x_{_q}, x_{_{\tilde{q}_j}}, 
x_{_{\tilde g}},x_{_{\chi_k^0}}, x_{_{\tilde{q}_i}})\Big[
{\bf Im}\Big((A_{_N}^q)_{_{kj}}({\cal Z}_{_{\tilde q}}^\dagger)_{_{j,1}}
\nonumber\\
&&\hspace{1.4cm}\times
(B_{_N}^q)_{_{ki}}({\cal Z}_{_{\tilde q}}^\dagger)_{_{i,1}}
e^{-i\theta_{_3}}\Big)-{\bf Im}\Big((B_{_N}^q)_{_{kj}}
({\cal Z}_{_{\tilde q}}^\dagger)_{_{j,2}}(A_{_N}^q)_{_{ki}}
({\cal Z}_{_{\tilde q}}^\dagger)_{_{i,2}}e^{i\theta_{_3}}\Big)\Big]
\nonumber\\
&&\hspace{1.4cm}
-m_{_{\chi_{_k}^0}}F_2(x_{_q}, x_{_{\tilde{q}_j}}, x_{_{\tilde g}},
x_{_{\chi_k^0}}, x_{_{\tilde{q}_i}})\Big[{\bf Im}
\Big((A_{_N}^q)_{_{kj}}({\cal Z}_{_{\tilde q}}^\dagger)_{_{j,2}}
(A_{_N}^q)_{_{ki}}({\cal Z}_{_{\tilde q}}^\dagger)_{_{i,1}}\Big)
\nonumber\\
&&\hspace{1.4cm}
-{\bf Im}\Big((B_{_N}^q)_{_{kj}}({\cal Z}_{_{\tilde q}}^\dagger)_{_{j,1}}
(B_{_N}^q)_{_{ki}}({\cal Z}_{_{\tilde q}}^\dagger)_{_{i,2}}\Big)\Big]\Big\}
\;,\nonumber\\
&&d_{_{\chi_{_k}^0(2)}}^g=
{4g_3\alpha\alpha_{_s}\over3(4\pi)^2s_{_{\rm w}}^2c_{_{\rm w}}^2m_{_{\rm w}}^2}
\Big\{|m_{_{\tilde g}}|F_4(x_{_q}, x_{_{\tilde{q}_j}}, x_{_{\tilde g}},
x_{_{\chi_k^0}}, x_{_{\tilde{q}_i}})\Big[{\bf Im}\Big((A_{_N}^q)_{_{kj}}
({\cal Z}_{_{\tilde q}}^\dagger)_{_{j,1}}
\nonumber\\
&&\hspace{1.4cm}\times
(B_{_N}^q)_{_{ki}}({\cal Z}_{_{\tilde q}}^\dagger)_{_{i,1}}
e^{-i\theta_{_3}}\Big)-{\bf Im}\Big((B_{_N}^q)_{_{kj}}
({\cal Z}_{_{\tilde q}}^\dagger)_{_{j,2}}(A_{_N}^q)_{_{ki}}
({\cal Z}_{_{\tilde q}}^\dagger)_{_{i,2}}e^{i\theta_{_3}}\Big)\Big]
\nonumber\\
&&\hspace{1.4cm}
-m_{_{\chi_{_k}^0}}F_5(x_{_q}, x_{_{\tilde{q}_j}}, x_{_{\tilde g}},
x_{_{\chi_k^0}}, x_{_{\tilde{q}_i}})\Big[{\bf Im}
\Big((A_{_N}^q)_{_{kj}}({\cal Z}_{_{\tilde q}}^\dagger)_{_{j,2}}
(A_{_N}^q)_{_{ki}}({\cal Z}_{_{\tilde q}}^\dagger)_{_{i,1}}\Big)
\nonumber\\
&&\hspace{1.4cm}
-{\bf Im}\Big((B_{_N}^q)_{_{kj}}({\cal Z}_{_{\tilde q}}^\dagger)_{_{j,1}}
(B_{_N}^q)_{_{ki}}({\cal Z}_{_{\tilde q}}^\dagger)_{_{i,2}}\Big)\Big]
\Big\}\;.
\label{eq11}
\end{eqnarray}

As for the two-loop gluino-chargino contributions to the c-quark
EDM and CEDM, we have
\begin{eqnarray}
&&d_{_{\chi_{_k}^\pm(2)}}^\gamma=
{4e\alpha\alpha_{_s}\over3(4\pi)^2s_{_{\rm w}}^2m_{_{\rm w}}^2}V_{_{qQ}}^\dagger V_{_{Qq}}
\Big\{|m_{_{\tilde g}}|F_6(x_{_Q}, x_{_{\tilde{Q}_j}}, x_{_{\tilde g}},
x_{_{\chi_k^\pm}}, x_{_{\tilde{q}_i}})\Big[{\bf Im}\Big((A_{_C}^Q)_{_{kj}}
({\cal Z}_{_{\tilde Q}}^\dagger)_{_{j,1}}
\nonumber\\
&&\hspace{1.4cm}\times
(B_{_C}^q)_{_{ki}}({\cal Z}_{_{\tilde q}}^\dagger)_{_{i,1}}
e^{-i\theta_{_3}}\Big)-{\bf Im}\Big((B_{_C}^Q)_{_{kj}}
({\cal Z}_{_{\tilde Q}}^\dagger)_{_{j,2}}
(A_{_C}^q)_{_{ki}}({\cal Z}_{_{\tilde q}}^\dagger)_{_{i,2}}
e^{i\theta_{_3}}\Big)\Big]
\nonumber\\
&&\hspace{1.4cm}
-m_{_{\chi_{_k}^\pm}}F_7(x_{_Q}, x_{_{\tilde{Q}_j}}, x_{_{\tilde g}},
x_{_{\chi_k^\pm}}, x_{_{\tilde{q}_i}})\Big[{\bf Im}
\Big((A_{_C}^Q)_{_{kj}}({\cal Z}_{_{\tilde Q}}^\dagger)_{_{j,2}}
(A_{_C}^q)_{_{ki}}({\cal Z}_{_{\tilde q}}^\dagger)_{_{i,1}}\Big)
\nonumber\\
&&\hspace{1.4cm}
-{\bf Im}\Big((B_{_C}^Q)_{_{kj}}({\cal Z}_{_{\tilde Q}}^\dagger)_{_{j,1}}
(B_{_C}^q)_{_{ki}}({\cal Z}_{_{\tilde q}}^\dagger)_{_{i,2}}\Big)\Big]\Big\}
\;,\nonumber\\
&&d_{_{\chi_{_k}^\pm(2)}}^g=
{4g_3\alpha\alpha_{_s}\over3(4\pi)^2s_{_{\rm w}}^2m_{_{\rm w}}^2}V_{_{qQ}}^\dagger V_{_{Qq}}
\Big\{|m_{_{\tilde g}}|F_4(x_{_Q}, x_{_{\tilde{Q}_j}}, x_{_{\tilde g}},
x_{_{\chi_k^\pm}}, x_{_{\tilde{q}_i}})\Big[{\bf Im}\Big((A_{_C}^Q)_{_{kj}}
({\cal Z}_{_{\tilde Q}}^\dagger)_{_{j,1}}
\nonumber\\
&&\hspace{1.4cm}\times
(B_{_C}^q)_{_{ki}}({\cal Z}_{_{\tilde q}}^\dagger)_{_{i,1}}
e^{-i\theta_{_3}}\Big)
-{\bf Im}\Big((B_{_C}^Q)_{_{kj}}({\cal Z}_{_{\tilde Q}}^\dagger)_{_{j,2}}
(A_{_C}^q)_{_{ki}}({\cal Z}_{_{\tilde q}}^\dagger)_{_{i,2}}
e^{i\theta_{_3}}\Big)\Big]
\nonumber\\
&&\hspace{1.4cm}
-m_{_{\chi_{_k}^\pm}}F_5(x_{_Q}, x_{_{\tilde{Q}_j}}, x_{_{\tilde g}},
x_{_{\chi_k^\pm}}, x_{_{\tilde{q}_i}})\Big[{\bf Im}
\Big((A_{_C}^Q)_{_{kj}}({\cal Z}_{_{\tilde Q}}^\dagger)_{_{j,2}}
(A_{_C}^q)_{_{ki}}({\cal Z}_{_{\tilde q}}^\dagger)_{_{i,1}}\Big)
\nonumber\\
&&\hspace{1.4cm}
-{\bf Im}\Big((B_{_C}^Q)_{_{kj}}({\cal Z}_{_{\tilde Q}}^\dagger)_{_{j,1}}
(B_{_C}^q)_{_{ki}}({\cal Z}_{_{\tilde q}}^\dagger)_{_{i,2}}\Big)\Big]
\Big\}\;.\nonumber\\
\label{eq12}
\end{eqnarray}
Note that the last terms of the $d_{_{\chi_{_k}^\pm(2)}}^\gamma$
and $d_{_{\chi_{_k}^\pm(2)}}^g$ are not proportional to the
suppression factor $m_{_q}/m_{_{\rm w}}$. This implies that the
two-loop gluino-chargino diagrams may be the dominant part of the
chargino contributions to the quark EDMs and CEDMs.

In order to account for resummation of the logarithmic
corrections, we should evolve the coefficients in the quark EDM
and CEDM operators  at the matching scale $\mu$ down to the charm
quark mass scale $m_{_c}\simeq1.2\;{\rm GeV}$ \cite{rge} using the
renormalization group equations (RGEs)
\begin{eqnarray}
&&d_{_q}^\gamma(\Lambda_{_\chi})=\eta_{_\gamma}d_{_q}^\gamma(\mu)
\;,\nonumber\\
&&d_{_q}^g(\Lambda_{_\chi})=\eta_{_g}d_{_q}^g(\mu)\;,
\label{eq14}
\end{eqnarray}
where $\eta_{_\gamma}\simeq1.53$ and $\eta_{_g}\simeq3.4$. At the
low energy scale, we need to include the contributions from the
quark CEDMs to the quark EDMs, when we evaluate the numerical
value of the charm quark EDM. This is realized by a naive
dimensional analysis \cite{nda} as
\begin{eqnarray}
&&d_{_c}=d_{_c}^\gamma+{e\over4\pi}d_{_c}^g\;.
\label{eq15}
\end{eqnarray}

\begin{figure}
\setlength{\unitlength}{1mm}
\begin{center}
\begin{picture}(0,140)(0,0)
\put(-80,-40){\includegraphics{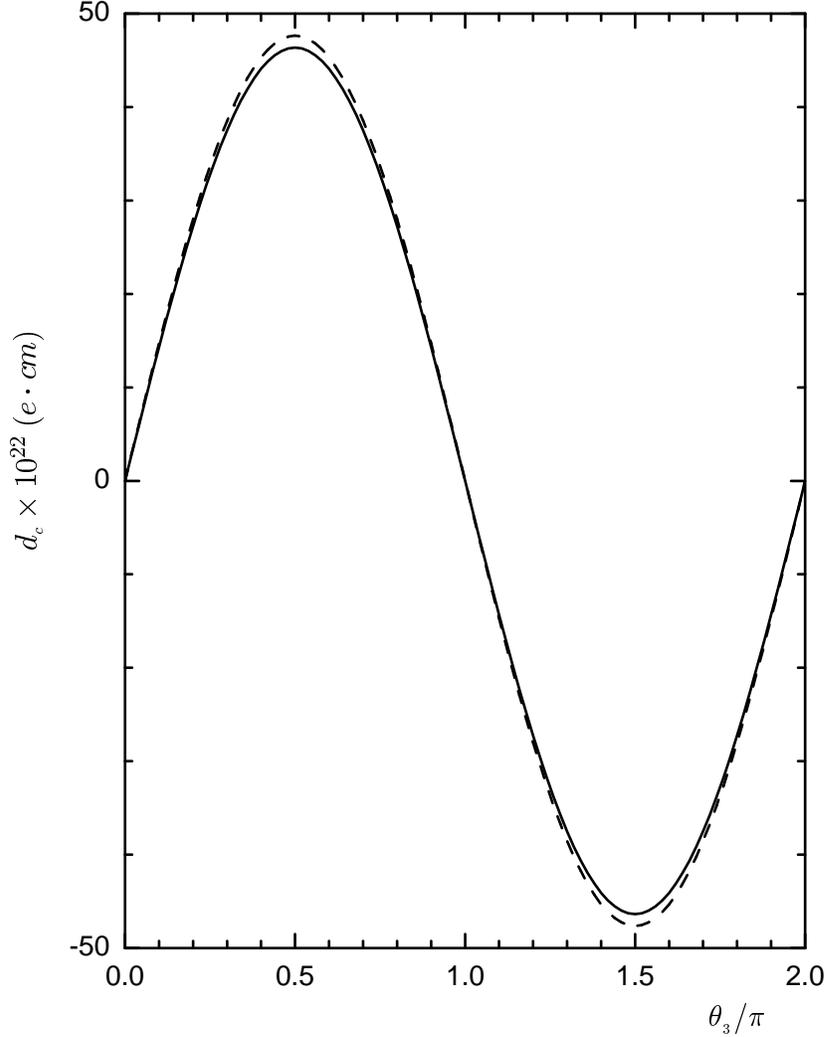}}
\end{picture}
\caption[]{The c-quark EDM varies with the $CP$ phase of $m_{_{\tilde g}}$
at $\tan\beta=2$, where the dashed line is the result at the
one-loop order, and the solid line is the result at the two-loop
order. The other parameters are taken as shown in text.}
\label{fig3}
\end{center}
\end{figure}
\begin{figure}
\setlength{\unitlength}{1mm}
\begin{center}
\begin{picture}(0,140)(0,0)
\put(-80,-40){\includegraphics{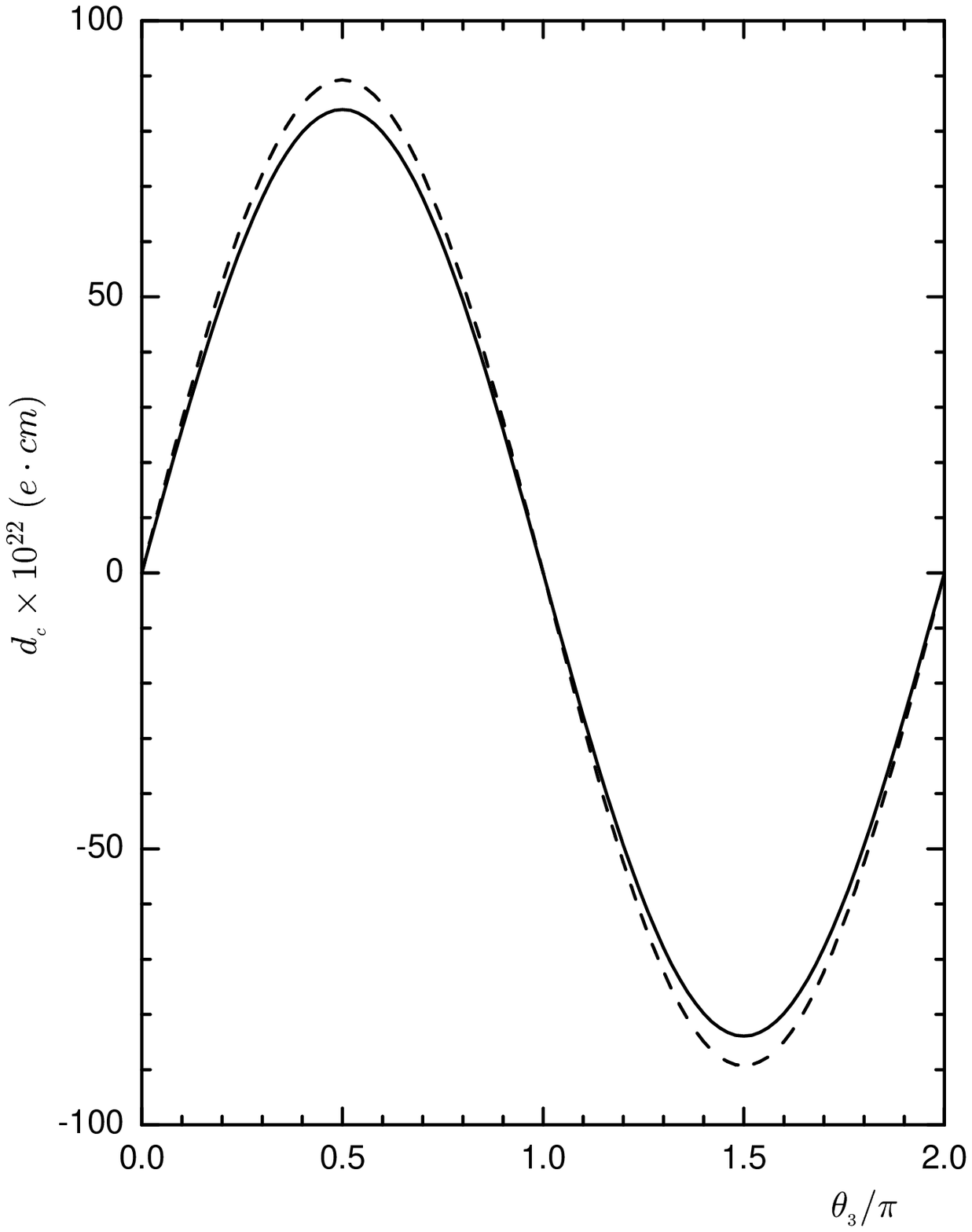}}
\end{picture}
\caption[]{The c-quark EDM varies with the $CP$ phase of $m_{_{\tilde g}}$
at $\tan\beta=20$, where the dashed line is the result at the
one-loop order, and the solid line is the result at the two-loop
order. The other parameters are taken as in text.} \label{fig4}
\end{center}
\end{figure}

\section{numerical result}

\subsection{In case of no flavor mixing}
\indent\indent In this subsection we present our numerical results
where no flavor mixing is introduced in the squark sector.
At present, the upper bound on the neutron EDM is
$d_{_n}\le1.1\times10^{-25}\;{e\cdot cm}$. In order to make
$d_{_n}$ consistent with this constraint, we take the squark
masses of the first generation as
$m_{_{\tilde{Q}_1}}=m_{_{\tilde{U}_1}}
=m_{_{\tilde{D}_1}}=20\;{\rm TeV}$, the $CP$ phase of the $\mu$
parameter $\theta_{_\mu}=0$. Without losing generality, we choose
the absolute values of the $\mu$ parameter and soft trilinear
quark couplings as $|\mu|={\bf A}_{_q} =300\;{\rm GeV}$ with
$q=u,\;d,\;\cdots,\;t$, and the soft $SU(2)\times U(1)$ gaugino
masses as $|m_{_1}|=|m_{_2}|=600\;{\rm GeV}$. For the soft masses
of the third generation squarks, we set
$m_{_{\tilde{Q}_3}}=m_{_{\tilde{U}_3}}
=m_{_{\tilde{D}_3}}=500\;{\rm GeV}$.

\begin{figure}
\setlength{\unitlength}{1mm}
\begin{center}
\begin{picture}(0,140)(0,0)
\put(-80,-40){\includegraphics{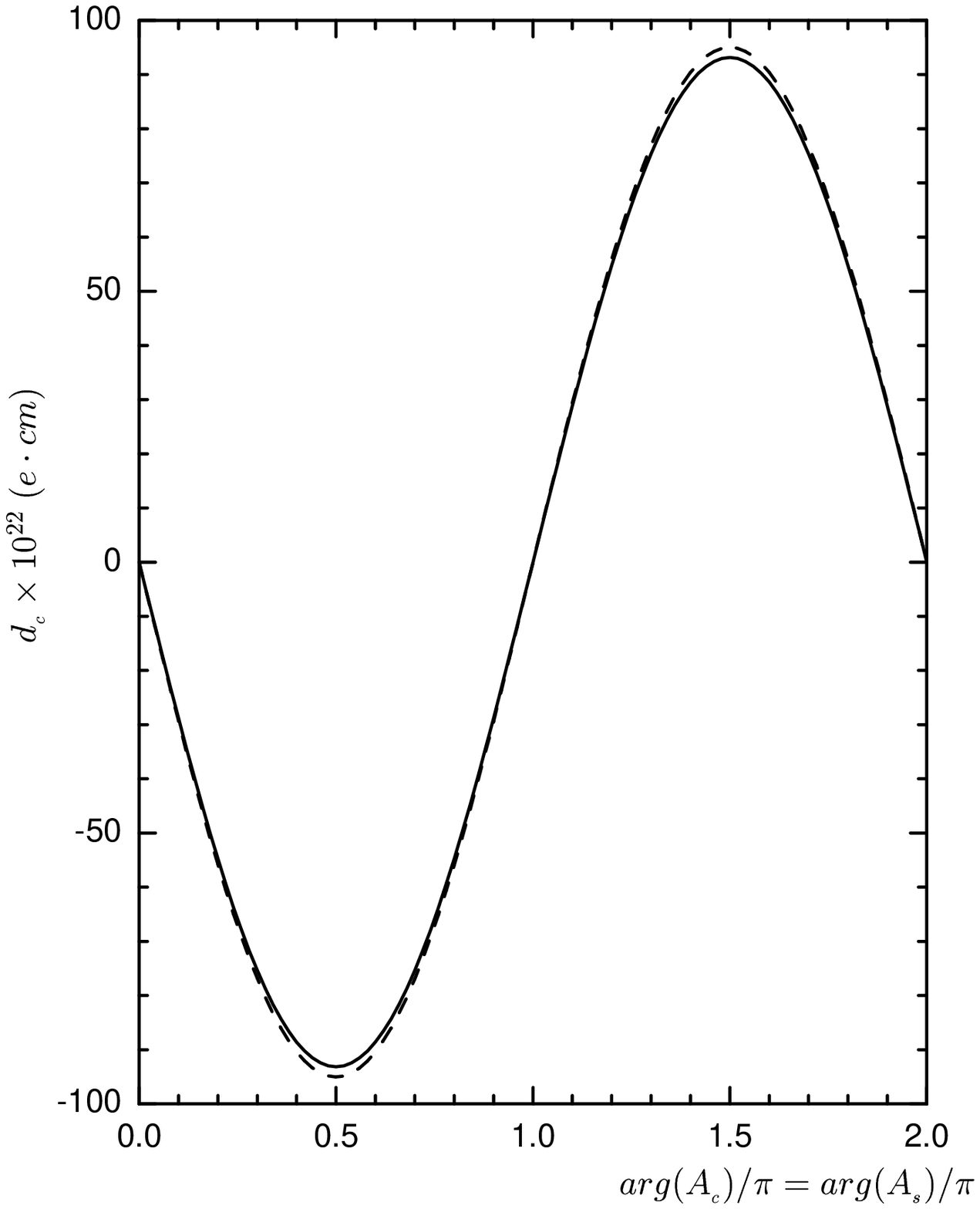}}
\end{picture}
\caption[]{The c-quark EDM varies with the $CP$ phases of soft
squark Yukawa couplings $arg({\bf A}_{_c})=arg({\bf A}_{_s})$ at
$\tan\beta=2$, where the dashed line is the result at  one-loop
order, and the solid line is the result at  two-loop order. The
other parameters are taken as in text.} \label{fig5}
\end{center}
\end{figure}
\begin{figure}
\setlength{\unitlength}{1mm}
\begin{center}
\begin{picture}(0,140)(0,0)
\put(-80,-40){\includegraphics{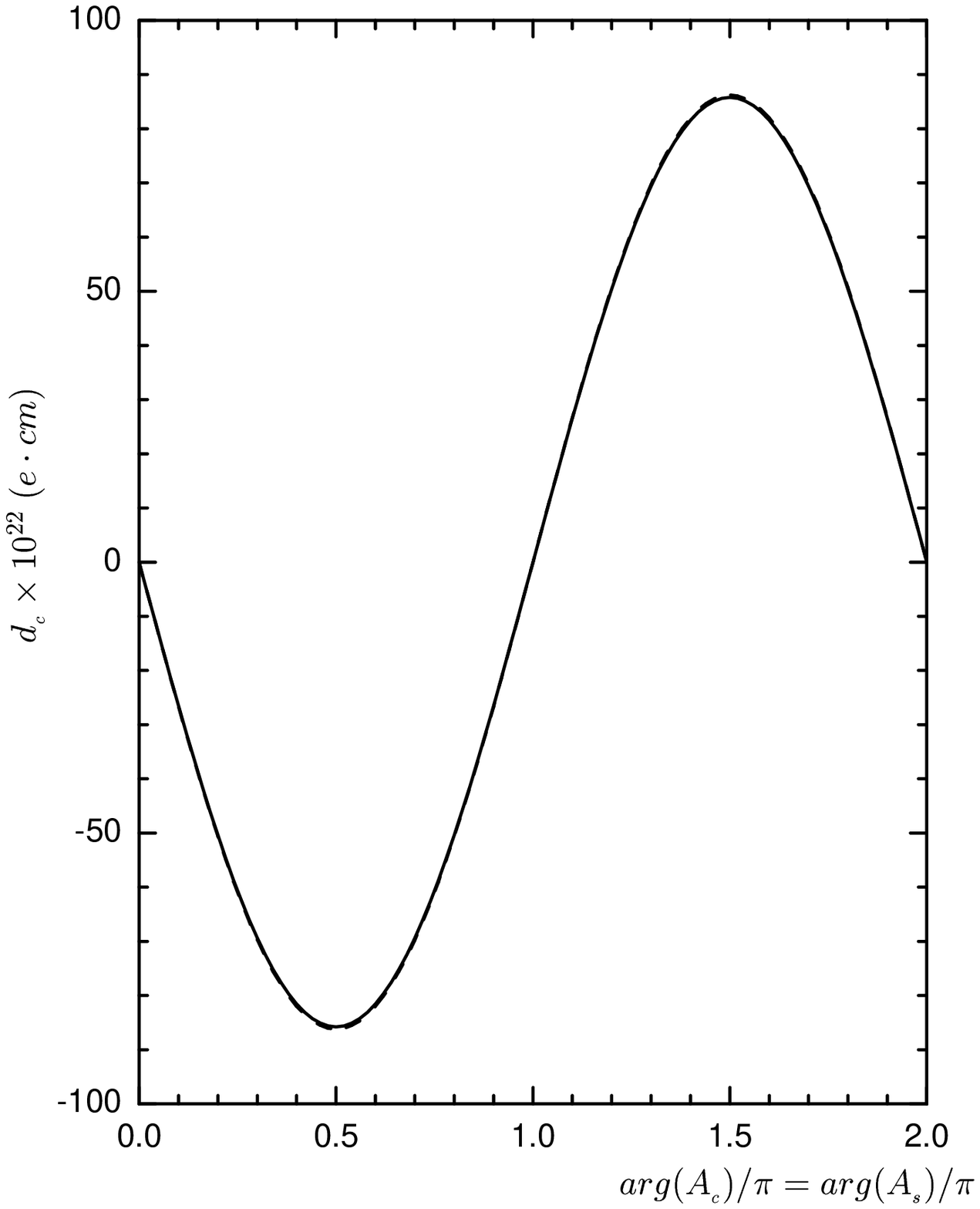}}
\end{picture}
\caption[]{The c-quark EDM varies with the $CP$ phases of soft
squark Yukawa couplings $arg({\bf A}_{_c})=arg({\bf A}_{_s})$ at
$\tan\beta=20$, where the dashed line is the result at one-loop
order, and the solid line is the result at two-loop order. The
other parameters are taken as in text.} \label{fig6}
\end{center}
\end{figure}

Taking $\tan\beta=2, arg({\bf A}_{_q})=0$, the $SU(3)$ gaugino
mass $|m_{_{\tilde g}}|=400\;{\rm GeV}$ and the soft masses  of the second
generation squarks as
$m_{_{\tilde{Q}_2}}=m_{_{\tilde{U}_2}}=m_{_{\tilde{D}_2}}=300\;{\rm
GeV}$, we plot the c-quark EDM $d_{_c}$ versus the $CP$ phase of
$m_{_{\tilde g}}$ in FIG. \ref{fig3}. With our choice of the parameter
space, the c-quark EDM $d_{_c}$ can reach $0.5\times
10^{-20}\;e\cdot cm$ while  the neutron EDM $d_{_n}$ is below the
experimental limit. Numerically, we find that the dominant
contributions to $d_{_c}$ originate from the gluino-squark sector
(including the one- and two-loop contributions). The sum of the
chargino-squark one-loop and chargino-gluino-squark two-loop
contributions to $d_{_c}$ is less than $10^{-24}\;e\cdot cm$.
Through the CKM mechanism, the contributions to $d_{_n}$ from the
second squarks (including squark-chargino one loop and
squark-chargino-gluino two-loop diagrams) is below
$10^{-26}\;e\cdot cm$. This fact can help us understanding why
$d_{_n}$ satisfies the experimental bound, while $d_{_c}$ can
reach a relatively large value. The choice of the parameter space
in FIG. \ref{fig4} is the same as that in FIG. \ref{fig3} except
for $\tan\beta=20$. Comparing with FIG. \ref{fig3}, the maximum of
$d_{_c}$ in FIG. \ref{fig3} is about $0.9\times10^{-20}\;e\cdot
cm$. Since the dependence of gluino-squark two-loop contributions
to the quark EDMs on the $CP$ phases is the same as that of
gluino-squark one-loop contributions to the quark EDMs on the $CP$
phases, the gluino-squark two-loop contributions to $d_{_c}$ is
below $10\%$ with our choice of the parameter space.

Now, we investigate how $d_{_c}$ varies with the $CP$ phases of
the soft squark Yukawa couplings. Taking $\tan\beta=2,
arg(m_{_{\tilde g}})=0$, we plot $d_{_c}$ versus the $CP$ phases $arg({\bf
A}_{_c})=arg({\bf A}_{_s})$ in FIG. \ref{fig5}, where the other
parameters are the same as in FIG. \ref{fig3}. It is clear that
the maximum of the absolute value of $d_{_c}$ is about
$0.9\times10^{-20}\;e\cdot cm$. With the same choice of the
parameter space as for FIG. \ref{fig5} except  $\tan\beta=20$,
$d_{_c}$ can reach $5\times10^{20}\;e\cdot cm$. Nevertheless, the
chargino sector which induces a contribution to $d_{_n}$ is above
the experimental upper limit. In order to suppress the chargino
contributions to $d_{_n}$, we choose
$m_{_{\tilde{Q}_2}}=m_{_{\tilde{U}_2}}=m_{_{\tilde{D}_2}}=500\;{\rm
GeV}$. Correspondingly, we set the $SU(3)$ gaugino mass as
$|m_{_{\tilde g}}|=600\;{\rm GeV}$. Within this scenario, the two-loop
gluino-squark contributions are much less than that of one-loop
gluino-squark contributions (FIG. \ref{fig6}).

In a similar way, we can investigate how $d_{_c}$ varies with the
$CP$ phases of the $SU(2)\times U(1)$ gaugino masses. However, the
contributions from the phases of the $SU(2)\times U(1)$ gaugino
masses  to $d_{_c}$ are below $10^{-23}\;e\cdot cm$ for our choice
of the parameter space.

\subsection{Effective SUSY}

In this subsection, we consider the case of the effective SUSY
scenario including mixing between the second and the third generation
squark. We fix the soft squark masses of the first two
generations as heavy as
$20 TeV$. 
We denote the different left-handed and right-handed squark masses
as $m_{\tilde{L}(\tilde{R})_{1,2,3}}$ respectively and
parameterize the the general form of the soft breaking mass matrix
as
\begin{equation}
m^2_{\tilde{Q}_{L,R}}=Z_{L,R} m_{\tilde{L},\tilde{R}}^2 Z_{L,R}^\dagger\ ,
\end{equation}
with
\begin{equation}
Z_L=\left( \begin{array}{ccc} 1& & \\ & \cos\theta_L & \sin\theta_L e^{i\phi_L/2}\\
 &  -\sin\theta_L e^{-i\phi_L/2} &\cos\theta_L \end{array}\right)\ ,\ \ \mbox{and}
\ \ \ m_{\tilde{L}}^2 =
\left( \begin{array}{ccc}m_{\tilde{L}_1}^2 &&\\& m_{\tilde{L}_2}^2 &
\\ && m_{\tilde{L}_3}^2 \end{array}\right)\ ,
\end{equation}
and the form for the right-handed sector is similar. The free
parameters are now $\tan\beta$, $m_{\tilde{L}_3}$,
$m_{\tilde{R}_3}$, $\theta_{L,R}$, $\phi_{L,R}$, and
$A=|A|e^{i\phi_A}$ for the squark sector; $\mu=|\mu|e^{i\phi_\mu}$
and $m_{1/2}$ for the chargino and neutralino sector; and the
gluino mass $m_{\tilde{g}}$. Assuming the gaugino masses are 
universal at the GUT scale, the phases of the gaugino masses can be rotated
away. Compared with the regular cases, the
effective SUSY scenario has four additional parameters:
$\theta_{L,R}$, $\phi_{L,R}$.

By our calculation, we are convinced that the dominant
contribution to $d_c$ comes from exchange of gluino. Therefore the
parameters related to the down squark sector and the chargino and
neutralino sector do not affect the numerical result
significantly. This fact helps us to reduce the parameter space.
We always take the same value of $m_{\tilde{L}_i}$, $\theta_{L,R}$
and $\phi_{L,R}$ for the up and down squark sector. We set
$\tan\beta=5$ and $\mu=150 GeV$ in our calculations.

\begin{figure}
\includegraphics[scale=0.45]{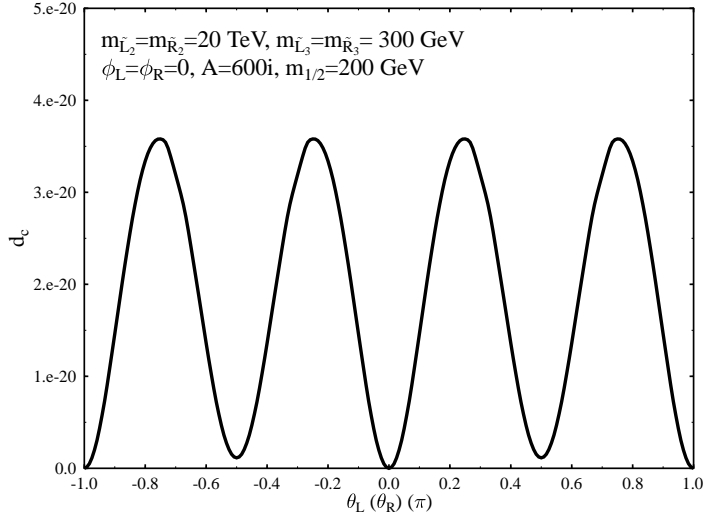}
\caption{\label{mix} $d_c$ as a function of
$\theta_{L}=\theta_{R}$ with $\tan\beta=5$, $\mu=150 GeV$,
$m_{\tilde{L}_2}=m_{\tilde{R}_2}=20 TeV$,
$m_{\tilde{L}_3}=m_{\tilde{R}_3}=300 GeV$, $m_{1/2}=200 GeV$,
$A=600i GeV$, $m_{\tilde{g}}=250 GeV$ and $\phi_L=\phi_R=0$. }
\end{figure}

By the above discussions we are left with only 8 free parameters:
$m_{\tilde{L}_3}$(=$m_{\tilde{R}_3}$), $\theta_{L,R}$,
$\phi_{L,R}$, and $A=|A|e^{i\phi_A}$. We first show the effects of
the mixing angle $\theta_{L,R}$ in Fig. \ref{mix}. We notice that
if the mixing angle is zero, $d_c$ is nearly zero, which is a
direct result of extremely heavy charm squark of about $20 TeV$.
For maximal mixing, $\theta_{L,R}=\pi/4$, we get the maximal
$d_c$.

\begin{figure}
\includegraphics[scale=0.45]{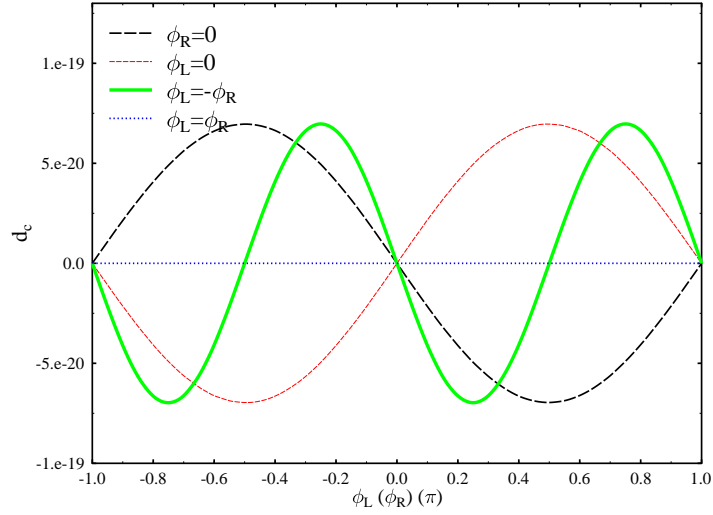}
\caption{\label{phi} $d_c$ as a function of $\theta_{L}$ (long
dashed line), $\theta_{R}$ (short dashed line),
$\theta_{L}=-\theta_{R}$ (solid line) and $\theta_{L}=\theta_{R}$
(dotted line) for $\theta_L=\theta_R=\pi/4$. Other parameters are
the same as those in Fig. \ref{mix}. }
\end{figure}

The contribution to $d_c$ from the phases of $\phi_{L,R}$ is
plotted in Fig. \ref{phi} for maximal mixing angle
$\theta_L=\theta_R=\pi/4$. It is interesting to notice that when
$\theta_{L}=\theta_{R}$ (dotted line) the contributions from
$\phi_{L}$ and $\phi_{R}$ cancel each other.

\begin{figure}
\includegraphics[scale=0.35]{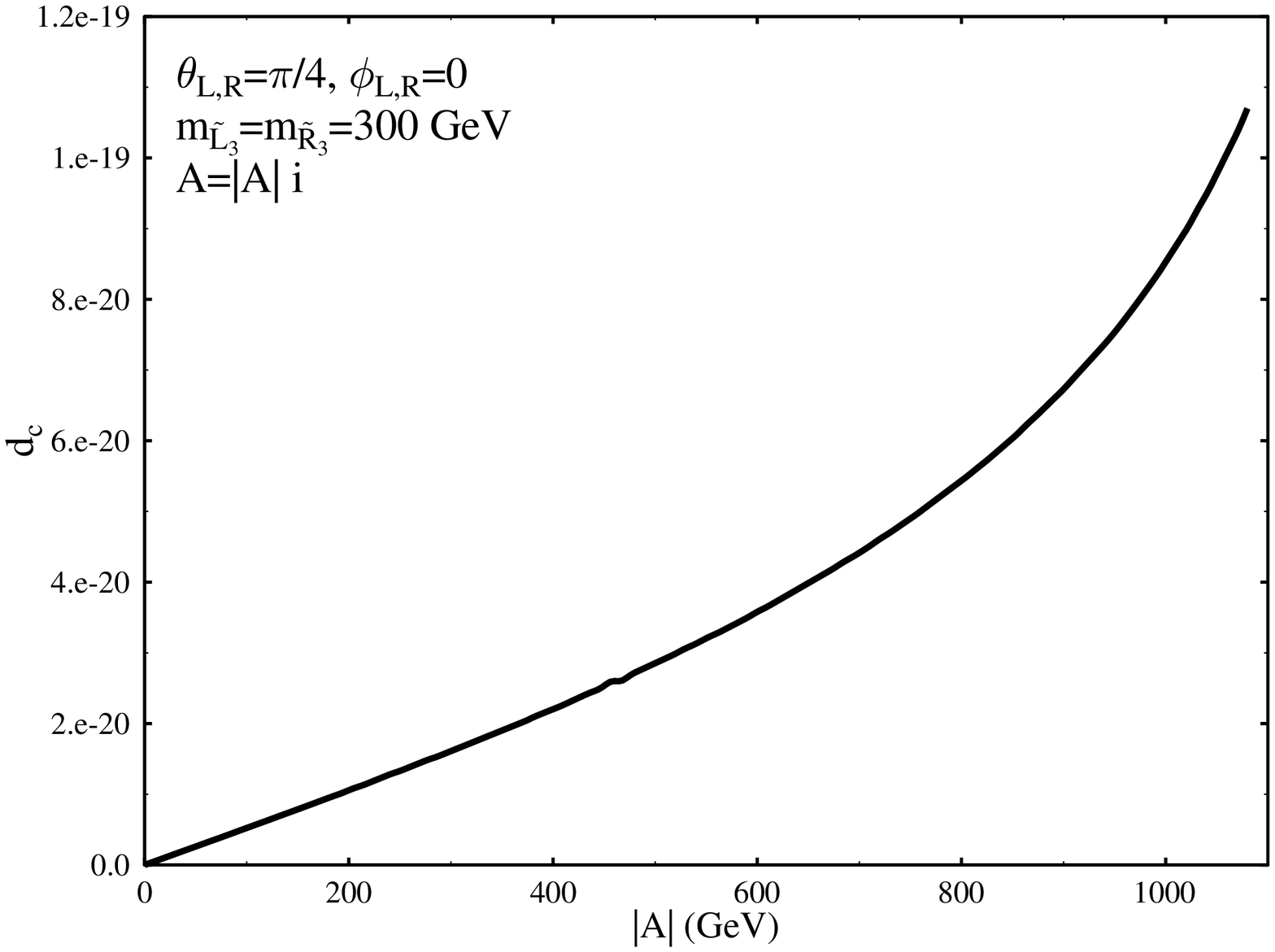}
\includegraphics[scale=0.35]{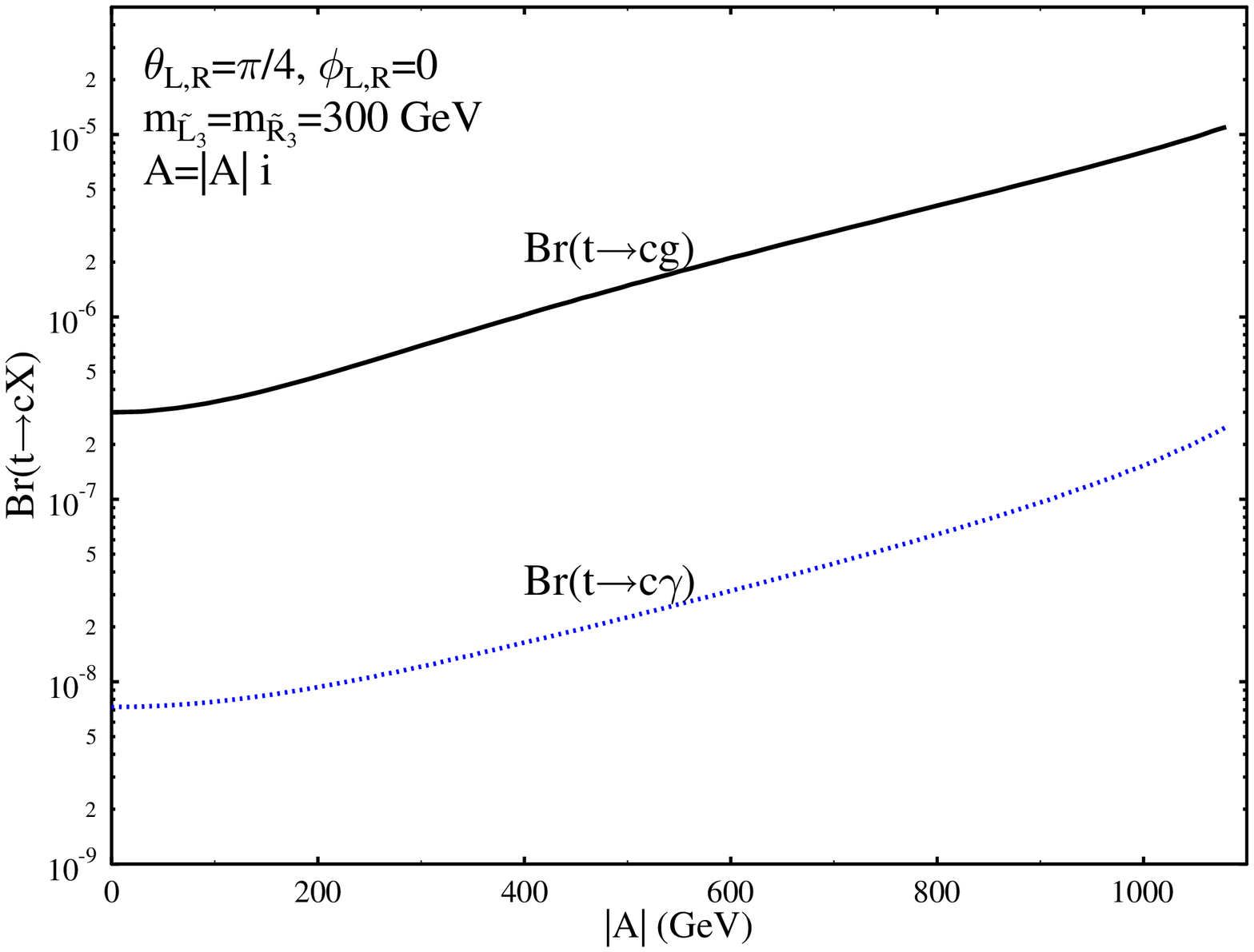}
\caption{\label{maga} $d_c$ (left panel) and the branching ratio
of $t\to c\gamma$ (right panel) as functions of $|A|$ for maximal
mixing angle $\theta_L=\theta_R=\pi/4$ and $\phi_{L,R}=0$. Other
parameters are the same as those in Fig. \ref{mix}. }
\end{figure}

The dependence of $d_c$ on the magnitude of $A$ is plotted in Fig.
\ref{maga} (on the left panel). For large $|A|$, not only the CP
violation is enhanced, but also the effects of the lighter stop
mass is suppressed. Therefore $d_c$ increases rapidly with
increasing value of $|A|$. The maximal value of $d_c$ can reach
about $\sim 10^{-19}e\cdot cm$.

The numerical results show that in the effective SUSY scenario,
where all the bounds set by the low energy experiments are
satisfied, the EDM of charm quark can be enhanced by an order of
magnitude due to the large mixing between the second and third
generations of up squark.

Finally, we show the flavor changing effects of large mixing
between scharm and stop. The branching ratios of $t\to c\gamma$
and $t\to c g$ induced by the mixing are plotted on the right
panel of Fig. \ref{maga}, which are below $\sim 10^{-5}$ and $\sim
10^{-7}$ respectively. We have adopted the LoopTools
package\cite{loop} to finish the loop integrations. In fact, the
present experiments do not constrain the mixing between scharm and
stop\cite{diaz}.

\section{$h_c$ production in $e^+e^-$ collider}

\begin{figure}
\includegraphics[scale=0.75]{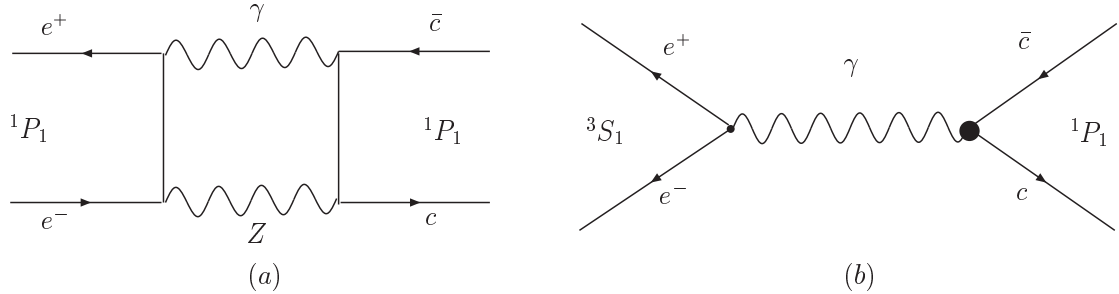}
\caption{\label{hcpro} The Feynman diagrams for production of the
$^1P_1$ charmonium resonance in $e^+e^-$ scattering from (a) the
CP-conserving and (b) CP-violating modes. The dark blob stands for
insertion of the effective operator (\ref{eq1}). }
\end{figure}

Recently the E835 Collaboration at Fermilab\cite{fermi}  
and the CLEO Collaboration\cite{cleoobs} 
have announced the observation of the $^1P_1$ CP-odd state
$h_c$. However, the direct production of the
$^1P_1$ charmonium resonance at the $e^+e^-$ annihilation plays an
important role in probing the charm quark EDM, since the coupling
of photon to the $^1P_1$ charmonium, as shown in Fig. \ref{hcpro}
(b), is identical to the effective operator in Eq. (\ref{eq1}).
The coupling of the current density\cite{cur}
\begin{equation}
\label{vect}
J_\mu(\bar{c}c|^1P_1)=\bar{c}(x)
\overleftrightarrow{{\partial}_\mu} \gamma_5
c(x)
\end{equation}
to a photon results in the effective operator (\ref{eq1}). We can
also check the quantum numbers of the process in Fig. \ref{hcpro}
(b). Transition from the initial $^3S_1$ ($J^{PC}=1^{--}$) state
to the final $^1P_1$ ($J^{PC}=1^{+-}$) state requires the
photon-charm-charm vertex, denoted as dark blob in Fig.
\ref{hcpro} (b), to be $-$ or $+$ under P and C transformations.
We find that the operator (\ref{eq1}) just satisfies the
requirement.
Under P and C transformations we have
\begin{eqnarray}
A_\mu&\stackrel{P}{\longrightarrow}& A^\mu,\ \ \
A_\mu\stackrel{C}{\longrightarrow} -A_\mu\ \ ,\\
\sigma_{\mu\nu}\gamma_5q^\nu&\stackrel{P}{\longrightarrow}& -\sigma^{\mu\nu}\gamma_5q_\nu,\ \
\sigma_{\mu\nu}\gamma_5q^\nu\stackrel{C}{\longrightarrow} -\sigma_{\mu\nu}\gamma_5q^\nu\ \ .
\end{eqnarray}
The operator (1) obviously possesses right quantum numbers.

In the SM, the $h_c$ ($1^1P_1$) meson can be produced via  $\gamma
Z$ and $ZZ$ box diagrams, as shown in Fig. \ref{hcpro} (a), which
are CP-conserving processes. As shown in Ref. \cite{bedm,cedm},
the effective Hamiltonian is given as
\begin{equation}
H_{SM}=\frac{\alpha}{3\pi\sqrt{2}}G_Fm_em_c\mathfrak{B}J_\mu(\bar{c}{c}|^1P_1)
\cdot J^\mu(e^+e^-|^1P_1)\ ,
\end{equation}
where the current $J_\mu$ is defined in Eq. (\ref{vect}) and
$\mathfrak{B}$ is a standard loop integral. In the minimal SUSY,
there is another diagram where $Z$ is replaced by a CP-odd Higgs
$A^0$, and generally it is considered to be smaller than the SM
contribution\cite{cedm}.

The effective Hamiltonian responsible for the CP-violating process
of $h_c$ production via the charm quark EDM, shown in Fig.
\ref{hcpro} (b), is given as
\begin{equation}
H_{SUSY}=\frac{4\pi\alpha}{M_{h_c}^2}\frac{d_c}{e}J_\mu(\bar{c}{c}|^1P_1)
\cdot J^\mu(e^+e^-|^3S_1)\ ,
\end{equation}
where $J^\mu(e^+e^-|^3S_1)=\bar{e}\gamma^\mu e$. Comparing the two
contributions we obtain the production amplitude of $h_c$ via the
CP-violating process and it can overtake the CP-conserving one if
the EDM of the charm quark exceeds the critical value\cite{cedm}
\begin{equation}
\left| \frac{d_c^{crit}}{e}\right|\sim \frac{G_Fm_e}{12\sqrt{2}\pi^2}
\frac{M_{h_c}^2}{M_Z^2}\log\frac{m_c}{m_e}\sim 10^{-26}cm\ .
\end{equation}
The numerical results presented in last section show that the EDM
of the charm quark in the effective SUSY is well above the
critical value $10^{-26}e\cdot cm$. Therefore, direct production of
$h_c$ at $e^+e^-$ annihilation is expected to be closely related
to the EDM of the charm quark or even determined by it.

Assuming that the center-of-mass energy of BEPC or CLEO-C is set at the
vicinity of $M_{h_c}(\sim 3524 MeV)$\cite{cleoobs}, we get
the cross section for $h_c$ production as \cite{bedm,cedm}
\begin{equation}
\sigma(e^+e^-\to h_c) = 27\left| \frac{d_c}{e} \right|^2 \left|
\frac{R'_P(0)}{R_S(0)}\right|^2 \sigma(e^+e^-\to ^3S_1)\sim 27
\left| \frac{M_{h_c}d_c}{e} \right|^2 \sigma(e^+e^-\to ^3S_1)\ ,
\end{equation}
where $R_P$ and $R_S$ are the wave functions of $h_c$ and $^3S_1$
states, respectively. The smallness of the EDM of charm quark
predicts a very small $h_c$ production rate, which is about 10
orders of magnitude smaller than $\sigma(e^+e^-\to ^3S_1)$ for
$h_c\sim 10^{-20}e\cdot cm$. With the upgraded BEPC\cite{bepc} and
CLEO-C programs\cite{cleo}, a data sample of $10^9\sim 10^{10}
J/\psi$ will be collected. We can estimate the number of $h_c$ events
as\cite{cedm}
\begin{equation}
N_{h_c}\sim 27\times N_{J/\psi} \left| \frac{M_{h_c}d_c}{e} \right|^2
\sim 10^{-10}\left|\frac{d_c/e}{10^{-20}cm} \right|^2\times N_{J/\psi}\ ,
\end{equation}
which predicts maximally $10^2\sim 10^3$ $h_c$ events per year in
the effective SUSY scenario with the upgraded BEPC and CLEO-C
luminosity. 
With the high energy resolution detector at BES-III\cite{bes},
the spectrum of gamma at the inclusive channel $h_c \to \gamma \eta_c$
can be well determined with high efficiency. 
The directly produced $h_c$ may be in the 
observable range at BES-III\cite{kuang}. 

\section{conclusion}

Considering the neutron EDM constraint, we investigate the EDM of
charm quark in the $CP$ violating MSSM. Typically, $d_{_c}$ can
reach about $10^{-20}\;e\cdot cm$ as we include the contributions
from c-quark CEDM. The mixing between the second and  third
generation squarks in the effective SUSY scenario further enhances
$d_{_c}$ to about $10^{-19}e\cdot cm$.

Large EDM of charm quark can induce a direct production of the
$1^1P_1$ CP-odd state charmonium, $h_c$, via a CP-violating
process at $e^+e^-$ annihilation. By the luminosity and efficiency
of the upgraded BEPC-II or CLEO-C (with $10^{10} J/\psi$
collected) if no $h_c$ is observed, a rigorous constraint on the
charm quark EDM would be set, as $d_c \lesssim 10^{-19} e\cdot cm$. On
the contrary, any signal of $h_c$ production will indicate large
EDM of the charm quark and becomes a clear evidence of new physics
beyond the SM.

The BEPC-II and CLEO-C will provide excellent opportunities to
probe many important theoretical objects. Among them, the charm
quark EDM is a prominent one because it is a clear signal of CP
violation and an evidence for new physics beyond the SM\cite{ma}.
Our numerical results indicate that the direct production rate of $h_c$ 
at the vicinity of $M_{h_c}\sim 3524MeV$\cite{cleoobs} may
be dominated by the charm quark EDM as long as new physics such as
the MSSM is involved. Observation of $h_c$ thus produced would clearly
indicate  new physics beyond the SM, conversely, a
negative result would set a rigorous constraint to the parameter
space for the new physics. Therefore we are looking forward to the
new data of BEPC-II and CLEO-C to help us gaining insight to the
new physics beyond the SM.

\begin{acknowledgments}
We thank X.Y. Shen and C.Z.Yuan for helpful discussions.
This work is supported in part by the National Natural Science
Foundation of China under the Grand No. 10105004, 10120130794, 19925523,
10047004, 10475042, 90303004, 
the Ministry of Science and Technology of China under
Grant NO. NKBRSF G19990754, and the Academy of Finland under the
contracts no.\ 104915 and 107293. 
\end{acknowledgments}

\end{document}